\documentstyle[aps,manuscript]{revtex}
\begin{document} 
\draft

\title{Nonlinear Micromechanical Casimir Oscillator}

\author{H. B. Chan\footnote{Electronic address: hochan@lucent.com}, 
V. A. Aksyuk, R. N. Kleiman, D. J. Bishop and 
Federico Capasso\footnote{Electronic address: fc@lucent.com}}
\address{Bell Laboratories, Lucent Technologies, Murray Hill, New Jersey 07974}

\maketitle
\date{Recieved}

\begin{abstract}
The Casimir force between uncharged metallic surfaces originates from 
quantum mechanical zero point fluctuations of the electromagnetic 
field. We demonstrate that this quantum electrodynamical effect has 
a profound influence on the oscillatory behavior of microstructures when surfaces 
are in close proximity ($\leq$ 100 $nm$). Frequency shifts, 
hysteretic behavior and bistability caused by the Casimir
force are observed in the frequency response 
of a periodically driven micromachined torsional oscillator. 
\end{abstract}

\narrowtext

Casimir forces are interactions between electrically neutral 
and highly conductive metals\cite{casimir,casrev}. 
They are regarded as one of the most striking manifestation 
of quantum fluctuations. The boundary conditions 
imposed on the electromagnetic fields lead to a spatial 
redistribution of the mode density with respect to free 
space, creating a spatial gradient of the zero point energy 
density and hence a net force between the metals.

The last few years have witnessed a resurgence of 
experiments\cite{mohideen,ederth,chan}
on these forces following the high precision measurements by 
Lamoreaux\cite{lamoreaux} using a torsional pendulum. 
Pioneering measurements were performed by 
Sparnaay\cite{sparnaay} followed later by the work of 
Van Blokland and Overbeek\cite{vanblokland} who accurately 
verified the existence of the Casimir effect. 
Between two parallel plates, the Casimir force is attractive 
and assumes the form $F_{c}=-\pi^{2}\hbar c A/240 z^{4}$, where $c$ is the 
speed of light, $\hbar$ is Planck constant/$2 \pi$, $A$ is the area of the 
plates and $z$ is their separation. 
In practice, one of interacting surfaces is usually chosen to be spherical 
to avoid alignment problems, modifying the force to 
$F_{cs}=-\pi^{3}\hbar c R/360 z^{3}$, where $R$ is the 
radius of the sphere\cite{derjaguin}. 

Casimir forces are inherently mesoscopic in nature since they can acquire 
substantial values when the separation between the metallic surfaces 
is reduced to $\leq$ 100 $nm$.  In addition, because of their topological 
nature associated with the dependence on the boundary conditions of 
the electromagnetic fields, their spatial dependence and sign can be 
controlled by tailoring the shapes of the interacting surfaces\cite{boyer}.
The above considerations have motivated us to investigate the effect of 
these quantum electrodynamical forces on the mechanical properties of 
artificial microstructures. MicroElectroMechanical Systems 
(MEMS)\cite{trimmer} are ideally suited for these studies because their 
moving parts can be engineered with high precision using state-of-the-art 
silicon integrated circuits technology and their separation can be 
controlled with high accuracy down to submicron distances\cite{serry}. 

In a previous paper we demonstrated the effect of the Casimir force 
on the {\it static} properties of micromechanical systems\cite{chan}. 
We used the deflection of a micromachined plate by a microsphere for a
high precision measurement of the Casimir force. Other studies have 
focused on adhesion and sticking of mobile parts 
in MEMS due to the Casimir effect\cite{buks}. 
In this letter we show that Casimir interactions have a profound 
effect on the {\it dynamic} properties of microstructures.  In particular 
we report on the experimental realization of a forced micromechanical 
nonlinear oscillator in which the anharmonic behavior arises solely 
from the Casimir effect.  A similar oscillator had been proposed 
and theoretically analyzed by Serry, Walliser and Maclay \cite{serry}. 
While there is a vast experimental literature on the hysteretic response 
and bistability of nonlinear oscillators in the context of quantum optics, 
solid-state physics, mechanics and electronics, the experiment described 
in this letter represents to our knowledge the first observation of 
bistability and hysteresis caused by a quantum electrodynamical effect.  

A simple model of the Casimir oscillator consists of a movable 
metallic plate subjected to the restoring force of a spring obeying 
Hooke's law and the nonlinear Casimir force arising from the interaction 
with a fixed metallic sphere (inset of Fig. 1). For separations $d$ 
larger than a critical value\cite{serry}, 
the system is bistable: the potential 
energy consists of a local minimum and a global minimum separated by 
a potential barrier (Fig. 1). The local minimum is a stable equilibrium 
position, about which the plate undergoes small oscillations. The 
Casimir force modifies the curvature of the confining potential 
around the minimum, thus changing the natural frequency of oscillation 
and also introduces high order terms in the potential, making the 
oscillations anharmonic. 

We realize such an oscillator making use of MEMS technology. The 
micromachined oscillator consists of a 3.5-$\mu m$-thick, 500-$\mu m^{2}$ 
polysilicon plate (metallized on the top with gold) free to rotate about 
two torsional rods on opposite edges (right inset in Fig. 2). The 
fabrication steps of a similar device used to study static effects of 
Casimir forces are described in Ref. \cite{chan}. We excite the torsional 
mode of oscillation by applying a driving voltage to one of the two 
electrodes that are fixed in position under the plate (left inset in 
Fig. 2). The driving voltage is a small ac excitation $V_{ac}$ with a dc 
bias $V_{dc1}$ to linearize the voltage dependence of the driving torque. 
The top plate is grounded while the detection electrode is connected 
to a dc voltage $V_{dc2}$ through a resistor. Oscillatory motion of the 
top plate leads to a time varying capacitance between the top plate 
and the detection electrode. For small oscillations, the change 
in capacitance is proportional to the rotation of the plate. The 
detection electrode is connected to an amplifier and a lock-in 
amplifier measures the output signal at the excitation frequency.

The measurement is performed at room temperature and at a pressure 
of less than 1 $mtorr$. Despite the soft torsional spring constant 
($k = 2.1 \times 10^{-8} Nm rad^{-1}$), the resonance frequency 
of the torsional mode is maintained reasonably high due to the small 
moment of inertia ($I = 7.1 \times 10^{-17} kg m^{2}$) of the top plate. 
The resonance peaks of the oscillator for different excitation voltages 
(Fig. 2) are fitted very well by the black curves representing driven 
motions of a damped harmonic oscillator. As expected, the resonance 
frequency remains constant at 2753.47 $Hz$ while the peak oscillation 
amplitude increases linearly with excitation. This clearly 
demonstrates that the oscillator behaves linearly in the absence of 
forces from external objects.

To investigate the effect of the Casimir force on the oscillator, 
we placed a 200-$\mu m$-diameter polystyrene sphere (metallized with gold) 
close to one side of the oscillator (inset in Fig. 3). The distance $z$ 
between the sphere and the equilibrium position of the top plate 
(i.e. without the periodic driving torque) is 
varied by a closed-loop piezoelectric stage. In the presence of 
the sphere, the equation of motion for the oscillator is given by:
\begin{equation}
    \ddot{\theta} + 2 \gamma \dot{\theta} + [\omega_{o}^{2}-(b^{2}/I)F'(z)]\theta 
= (\tau/I) cos \omega t -\alpha \theta^{2} - \beta \theta^{3}
\label{eq:eqm}
\end{equation}
\noindent where $\tau$ is the amplitude of the driving torque, $b$ is the 
lateral distance of the sphere from the center of the top plate,
$\omega_{o}=\sqrt{k/I}$ is the fundamental frequency of the oscillator,
$\gamma$ is the damping coefficient, $\alpha = b^{3}F''(z)/2I$ and
$\beta = -b^{4}F'''(z)/6I$. $F'(z)$, $F''(z)$ and $F'''(z)$ denote the first, 
second and third spatial derivative of the external force $F$ respectively, 
evaluated at distance $z$. In our experiment, $F$ is either 
the Casimir force or an applied electrostatic force between the sphere 
and the top plate. To obtain Eq. (\ref{eq:eqm}), $F(z-b \theta)$ has
been Taylor expanded about $z$ up to $\theta ^{3}$. For small oscillations 
where the nonlinear terms $\theta^{2}$ and $\theta^{3}$ can be neglected, 
the external force modifies the mechanical resonance frequency of the 
oscillator by an amount that is proportional to the force gradient:
\begin{equation}
    \omega_{1} = \omega_{o}[1-b^{2}F'(z)/2I\omega_{o}^{2}]
\label{eq:w1}
\end{equation}

To calculate $b$, we deliberately apply a voltage to the sphere to set up 
an electrostatic force gradient. Then we record the change in resonance 
frequency of the oscillator as we vary the distance $z$ by changing the 
piezo extension. The gradient of the electrostatic force $F_{e}$ between 
the sphere and the top plate is given by:
\begin{equation}
    F_{e}'(z) = \epsilon \pi R (V-V_{o})^{2}/(\Delta z+z_{o})^2 $ for $ \Delta z+z_{o}\ll R
\label{eq:fe}
\end{equation}
\noindent where $\epsilon_{o}$ is the permittivity of vacuum, $R$ is the 
radius of the sphere, $V$ is the voltage applied to the sphere, $V_{o}$ is 
the residual voltage on the sphere, $z_{o}$ is the distance of closest 
approach of the sphere from the plate and $\Delta z$ is the separation between the 
sphere and plate measured from $z_{o}$, so that $z=\Delta z+z_{o}$. 
Here $z_{o}$ is not the minimum achievable 
separation. While it is possible to extend the piezo further to decrease the value 
of $z_{o}$, we did not attempt to do so in order to prevent the top plate of the 
oscillator from jumping into contact with the sphere. The residual voltage $V_{o}$ 
arises from the work function difference of the two gold surfaces 
(on the sphere and on the top plate) as a result of slight variations in the 
preparation of the films\cite{chan,vanblokland}. 
$V_{o}$ is found to be 75 $mV$ by fixing the distance $z$ 
and identifying the voltage $V$ at which the maximum of the quadratic voltage 
dependence of the resonance frequency occurs.

We perform a fit of the resonance frequency shift of the oscillator 
(solid line in Fig. 3) in response to the electrostatic force gradient using 
Eqs. (\ref{eq:w1}) and (\ref{eq:fe}) with $z_{o}$ and $b$ 
as fitting parameters (determined to be 122.4 $nm$ and 
131.0 $\mu m$ respectively). We then set the voltage $V$ on the sphere equal to 
the residual voltage $V_{o}$ to eliminate the electrostatic contributions to 
the force gradient. In Fig. 3, the squares are the shifts in 
resonance frequency obtained when we repeat the measurement with $V = V_{o}$. 
The dash line is a fit to the predicted frequency shift (Eqs. (\ref{eq:w1}))
due to the Casimir force gradient assuming perfectly conducting surfaces:
\begin{equation}
    F_{cs}'(z) = \pi^{3} \hbar c R /120 (\Delta z+z_{1})^4 
\label{eq:fc}
\end{equation}
\noindent where $z_{1}$ is the distance of the sphere from the plate at 
the distance of closest approach (determined to be 85.9 $nm$ from the fit). 
As we discussed in an earlier experiment\cite{chan}, for an exact comparison 
of data with theory the finite conductivity and surface roughness of the metal 
films must be taken into account\cite{corrections}. 
However, we do not attempt such a precise comparison here because 
contributions from the higher order derivatives of the Casimir force 
modify the shift calculated from Eq. (\ref{eq:w1}) by more than 5\% 
at the smallest separations, as we discuss later. As a result the 
uncertainty in deducing the Casimir force from the measured frequency shift 
(using Eq. (\ref{eq:w1})) is in excess of 5\% at the closest separation. 
This value is significantly larger than the experimental error in our 
earlier static measurement ($\leq$ 1\%). However, in the linear regime 
dynamic techniques are expected to ultimately yield a higher sensitivity 
than static measurements\cite{bressi}.

To demonstrate the nonlinear effects introduced by the Casimir force, we first 
retract the piezo until the sphere is more than 3.3 $\mu m$ away from the oscillating 
plate so that the Casimir force has negligible effect on the oscillations. The 
measured frequency response shows a resonance peak that is characteristic of a 
driven harmonic oscillator (peak I in Fig 4a), regardless of whether the 
frequency is swept up (hollow squares) or down (solid circles). This ensures the 
excitation voltage is small enough so that intrinsic nonlinear effects in the 
oscillator are negligible in the absence of the Casimir force. We then extend the piezo 
to bring the sphere close to the top plate while maintaining the excitation 
voltage at fixed amplitude. The resonance peak shifts to lower frequencies 
(peaks II, III and IV), by an amount that is consistent with the distance 
dependence in Fig. 3. Moreover, the shape of the resonance peak deviates from 
that of a driven harmonic oscillator and becomes asymmetric. As the distance 
decreases, the asymmetry becomes stronger and hysteresis occurs. This 
reproducible hysteretic behavior is characteristic of strongly nonlinear 
oscillations\cite{landau}. For a given excitation $\tau$ and frequency $\omega$, 
the amplitude of oscillation $A$ is given by the roots of the following equation:
\begin{equation}
    A^{2}[(\omega-\omega_{1}-\kappa A^{2})^{2}+ \lambda^{2}] = \tau^{2}/4I^{2}\omega_{1}^{2}
\label{eq:nl}
\end{equation}
\noindent where $\kappa = 3\beta/8 \omega_{1}-5 \alpha^{2}/12 \omega_{1}^{3}$ 
characterizes the non-linear effects. When the non-linearity is weak, Eq. (\ref{eq:nl}) 
only has a single positive solution for $A^{2}$. In the presence of strong 
non-linearity, such as those introduced by the Casimir force in our experiment, 
the oscillation amplitude $A$ becomes triple valued for a range of frequency, 
corresponding to the 3 positive roots of $A^{2}$ in Eq. (\ref{eq:nl}). The solid lines in 
Fig. 4a show the predicted frequency response of the oscillator
with $w_{1}$ and $\kappa$ determined by the first, second and third spatial 
derivatives of the Casimir force at $z$ = 98 $nm$, 
116.5 $nm$, 141 $nm$ and 3.3 $\mu m$ respectively. 
The values of other parameters ($\gamma$, $b$, $I$, $\tau$) are identical for all four 
resonance peaks. At a particular distance, the spatial derivatives of the 
Casimir force determine both the frequency and the shape of the resonance 
peaks without any other adjustable parameters. Indeed, the shape and the 
frequency of peaks II and III agree well with Eq. (\ref{eq:nl}). For peak IV, the 
hysteretic effects are very strong and deviations from Eq. (\ref{eq:nl}) become apparent. 
This discrepancy arises from contributions of higher order spatial derivatives 
that we neglected in the series expansion of the Casimir force (Eq. (\ref{eq:eqm})), 
as well as corrections to the Casimir force as a result of finite conductivity 
and roughness of the surfaces\cite{corrections}.

An alternative way to demonstrate the ``memory'' effect of the oscillator is to 
maintain the excitation at a fixed frequency and vary the distance between 
the sphere and the plate (Fig. 4b). As the distance
changes, the resonance frequency $w_{1}$ of the oscillator 
shifts, to first order because of the changing force gradient (Eq. (\ref{eq:w1})). 
In region 1, the fixed excitation frequency is higher than the resonance 
frequency and vice versa for region 3. In region 2, the amplitude of 
oscillation depends on the history of the plate position. Depending on 
whether the plate was in region 1 or region 3 before it enters region 2, 
the amplitude of oscillation differs by up to a factor of 6. This oscillator 
therefore acts as a sensor for the separation between the two surfaces.

In Fig. 4a, we used a constant quality factor $Q$ = 7150 to fit all four 
resonance peaks at different distances. Further improvements in sensitivity 
could enable us to explore possible changes in $Q$ with distance. 
There has been an interesting prediction\cite{levitov} that 
dissipative retarded van der Waals forces
can arise between surfaces in relative motion due to the exchange of virtual 
photons which couple to acoustic phonons in the material. Similar dissipative 
Casimir forces can arise between metals; here virtual photons would 
couple to particle-hole excitations in the metal\cite{levitov2}. 
This would lead to changes in the $Q$ of our oscillator with position. 
It is also interesting to point out that the non-uniform relative 
acceleration of the metal and the sphere will lead, at least in principle, 
to an additional damping mechanism associated with the parametric 
down-conversion of vibrational quanta into pairs of photons, 
a quantum electrodynamical effect associated with the nonlinear 
properties of vacuum. This phenomenon, which was investigated 
theoretically by Lambrecht and Reynaud in the context of a vibrating 
parallel plate capacitor\cite{lambrecht}, is an example of the so called dynamical 
Casimir effect, i.e. the non-thermal radiation emitted by dielectric bodies 
in a state of non-uniform acceleration\cite{fulling}. Although this effect 
is completely negligible in our system, it does represent a fundamental lower limit to 
the damping of the Casimir oscillator. 

Finally, we remark that nonlinear effects in a mechanical oscillator 
arising from ordinary electrostatic forces were studied by several 
groups\cite{buks2,krommer}. In particular, Buks {\it et al}.\cite{buks2} 
considered the role of the Casimir force in such nonlinear oscillators. 
While the relative strength of the electrostatic force to the Casimir 
force was not given, using their smallest 
separation of 0.77 $\mu m$ and 30 $V$ between the surfaces we
estimate that in their experiment the Casimir force is roughly
$10^{6}$ times smaller than the electrostatic force before pull-in
assuming a simple parallel plate model. Therefore, quantum effects such 
as the Casimir force have negligible effect on the non-linearity observed 
in their oscillator, though the pull-in and sticking of their oscillator 
might in part be due to the Casimir force.

We thank L. S. Levitov, M. Schaden, L. Spruch, R. Onofrio, 
M. R. Andrews, D. Abusch-Magder, 
R. de Picciotto, M. I. Dykman, C. F. Gmachl, A. 
Moustakas, L. N. Pfeiffer, P. M. Platzman and N. Zhitenev for 
assistance and useful discussions.

\begin{figure}

\caption{Inset: A simple model of the nonlinear Casimir oscillator 
(not to scale). Main figure: 
Elastic potential energy of the spring (dotted line, spring constant = 
0.019 $Nm^{-1}$), energy associated with the Casimir attraction 
(dash line) and total potential energy (solid line) as a function of 
plate displacement. The distance $d$, measured between the 
sphere (100 $\mu m$ radius) and the equilibrium position of the 
plate in the absence of the Casimir force, is chosen to be 
40 $nm$.}

\end{figure}

\begin{figure}

\caption{Resonance peaks of the torsional oscillator at excitation 
voltage amplitudes of 35.4 $\mu V$ (triangles) and 72.5 $\mu V$ (circles). 
The solid lines are 
fits to the data based on a driven harmonic oscillator. Inset 
(right): Schematic of the torsional oscillator (not to scale). 
Inset (left): Cross-section of the device with the electrical 
connections and measurement circuit.}

\end{figure}

\begin{figure}

\caption{Change in resonance frequency of the oscillator 
in response to the electrostatic force (circles, $V$ = 408.5 $mV$) 
and Casimir 
force (squares) as a function of distance. The amplitude of 
the excitation is 8.2 $\mu V$, 
producing oscillations of the plate with amplitude of 5.8 $nm$ 
at its closest point to the sphere. The 
solid and dash lines are fits obtained with Eqs. (\ref{eq:fe}) 
and (\ref{eq:fc})
respectively. (Inset) Schematic of the experiment 
(not to scale).
The oscillation angle $\theta$ indicated by the curved arrow 
is measured from the equilibrium position of the plate in the 
absence of driving torque.}

\end{figure}

\begin{figure}

\caption{(a) Hysteresis in the frequency response induced by the 
Casimir force on an otherwise linear oscillator. 
Hollow squares (solid circles) 
are recorded with increasing (decreasing) frequency. 
The distance $z$ 
between the oscillator and the sphere is 3.3 $\mu m$, 141 $nm$, 
116.5 $nm$ and 98 $nm$ for peaks I, II, III and IV respectively. 
The excitation amplitude is maintained constant at 55.5 $\mu V$ 
for all 4 separations. The solid lines are the calculated response 
using Eq. (\ref{eq:nl}), with $\kappa = 0$, $-3.1 \times 10^{7}$,  
$-1.0 \times 10^{8}$ and 
$-2.8 \times 10^{8}$ $rad^{-2} s^{-1}$ for peaks I, II, III and 
IV respectively. The peak oscillation amplitude for the plate is 
39 $nm$ at its closest point to the sphere. (b) Oscillation amplitude 
as a function of distance with excitation frequency fixed at 
2748 $Hz$.}

\end{figure}

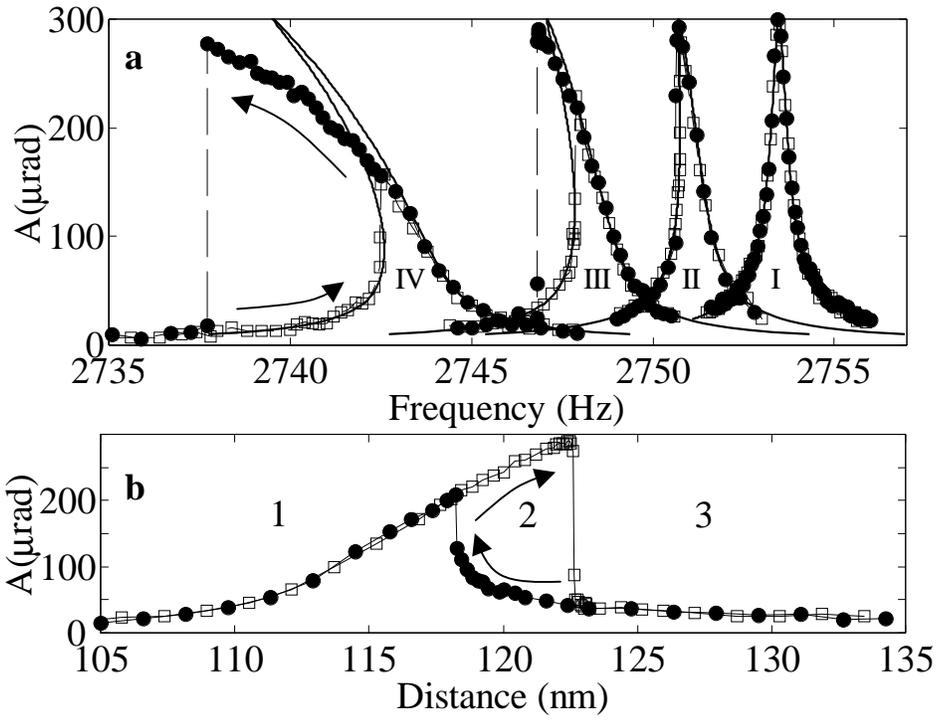
\end{document}